# Comment on "Crystallite size dependent exchange bias in MgFe$_2$O$_4$ thin films on Si (100)", [J. Appl. Phys. **124**, 053901 (2018)]

Himadri Roy Dakua

Department of physics, Indian Institute of Technology Bombay, Mumbai, India – 400076

**Abstract**

K. Mallick and P. S. A. Kumar[1] had reported exchange bias effect in Mg-ferrite thin films, deposited on Si substrate (with a buffer layer of MgO) using Pulsed Laser Deposition (PLD) technique. The authors had presented the temperature dependence exchange bias effect, field dependence exchange bias effect and training effect of a selected Magnesium ferrite thin film of thickness 132 nm. These studies were followed by the film thickness dependence of exchange bias effect. However, the data presented for the 132 nm thick film shows mutually contradicting values in each and every figures. Here, I point out these highly self-contradicting data in this comment.

**Comment**

Exchange Bias (EB) effect is a well-known phenomenon within the society, working in magnetism and magnetic materials. The EB effect is generally characterized by a horizontal shift (along the field axis) in the magnetic hysteresis loop of a Field Cooled (FC) system.[2, 3] Fig. 1 shows schematic diagram of Zero Field Cooled (ZFC) and FC M-H loops of a typical exchange bias system. The exchange bias field ($H_E$) and the coercivity ($H_C$) is calculated as - $H_E = (H_1+H_2)/2$ and $H_C = (H_1 - H_2)/2$, where $H_1$ and $H_2$ are two coercive fields shown in Fig. 1. Mallick and Kumar also used similar formulae to calculate the magnitude of $H_E$ and the coercivity ($H_C$) of their films.[1] Since the FC M-H loop shifted along the negative field axis, it must follow the simple relation $|H_2| > |H_1|$ and $|H_2| > H_C$.[4] In training effect, the magnitude of $H_E$ gradually decreases as the $|H_2|$ and $|H_1|$ tend to be equal with increasing M-H loop iterations.[5]

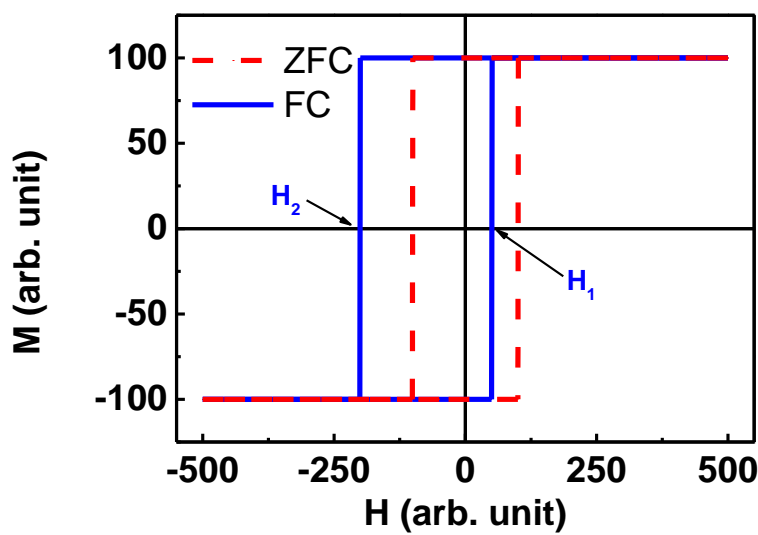

Fig. 1. Schematic diagram of ZFC and FC M-H loops of a typical exchange bias system

Mallick and Kumar had presented the coercivity ($H_C$) data of a 132 nm thick Mg-ferrite film in different figures of their paper.[1] However, the coercivity values of the film are found to be different in different figures for an identical measurement. Here I have pointed out these differences one by one.

1. The inset of FIG 1 (c) of the said paper shows the expanded view of + 6 kOe and - 6 kOe FC M-H loops of the 132 nm thick film measured at 10 K. Here, the values of $|H_1|$ and $|H_2|$ of the + 6 kOe FC M-H loop are ~ 400 Oe and ~750 Oe respectively. Therefore, the coercivity ($H_C$) of + 6 kOe FC M-H loop of the film should be $H_C$ = ~ 575 Oe at 10 K.

2. In FIG. 2 (a), the authors had presented the cooling field dependence of $H_C$ (at 10 K) of the same 132 nm thick film. Here the estimated value of $H_C$ of 6 kOe FC M-H loop is within the range 400 < $H_C$ < 460 Oe, which is very small than the estimated value reported in FIG. 1 (c).

3. While in FIG. 2 (b), the $H_C$ at 10 K of the 6 kOe FC M-H loop of same 132 nm thick film is reported as ~ 360 Oe!

4. Mallik and kumar had also presented the training effect of the same film at 5 K after field cooling in 20 kOe magnetic field. They presented the negative descending branches of the M-H loops in FIG 4 (a). The value of $|H_2|$ of the first M-H loop iteration is ~ 205 Oe. Since the 20 kOe FC M-H loop shifted along negative field axis, so one must find $|H_2| > |H_1|$ and $|H_2| > H_C$. However, the authors reported (in FIG. 4 (b)) $H_C = \sim 325$ Oe for the first iteration which is much higher than the $|H_2|$ (~205 Oe)!

Such huge discrepancies in the coercivity values of the same film in different figures clearly tell that the exchange bias field might be also altered from the real values and true nature. Therefore the discussion and conclusion based on these results could not be reliable.